\documentclass[twocolumn]{jpsj2} %% two-column layout
\def\stacksymbols #1#2#3#4{\def\theguybelow{#2}
    \def\verticalposition{\lower#3pt}
    \def\spacingwithinsymbol{\baselineskip0pt\lineskip#4pt}
    \mathrel{\mathpalette\intermediary#1}}
\def\intermediary#1#2{\verticalposition\vbox{\spacingwithinsymbol
      \everycr={}\tabskip0pt
      \halign{$\mathsurround0pt#1\hfil##\hfil$\crcr#2\crcr
               \theguybelow\crcr}}}

\title{
Spatial Structures of Anomalously Localized States in Tail Regions at the Anderson Transition}
\author{H. Obuse and K. Yakubo}

\inst{Department of Applied Physics,
Graduate School of Engineering, Hokkaido University, Sapporo
060-8628, Japan.}

\abst{ We study spatial structures of anomalously localized states
(ALS) in tail regions at the critical point of the Anderson transition
in the two-dimensional symplectic class. In order to examine tail
structures of ALS, we apply the multifractal analysis only for the tail
region of ALS and compare with the whole structure. It is found
that the amplitude distribution in the tail region of ALS is
multifractal and values of exponents characterizing multifractality
are the same with those for typical multifractal wavefunctions in
this universality class.}

\kword{Anderson transition, critical states, multifractal,
anomalously localized states, two-dimensional symplectic system}

\begin{document}

\maketitle

\section{Introduction}
\label{sec:1}

Disorder induced metal-insulator transition at absolute zero
temperature, namely, the Anderson transition, has been extensively
studied since the seminal paper by Anderson.\cite{Anderson1} Although
the vast amount of knowledge on this physical phenomenon has been
accumulated for half a century, there still exist fundamental problems
on properties of non-interacting electrons in disordered systems. One
of such unsettled questions is anomalously localized states
(ALS)\cite{Altshuler1,Mirlin1} in which most of amplitudes concentrate
on a narrow spatial region even in a metallic phase. The existence of
ALS was analytically predicted, and the nature of ALS has been
numerically and analytically studied.
\cite{Muzykantskii1,Falko1,Smolyarenko1,Kogan1,Uski1,Nikolic1,Nikolic2,Uski2,Nikolic3,Apalkov1,Apalkov2,Kottos1,Kottos2,Apalkov3}
Most of previous works on ALS focused on the amplitude distribution
function, which mainly reflects peak structures of ALS. Falko
{\it et al.}\cite{Falko1} and Uski {\it et al.}\cite{Uski2} have
predicted that the spatial correlation function of amplitudes
has a long tail. This implies that ALS are not truely localized. The
tail structure of ALS is crucial for understanding transport
properties of metals.

Recently, we have shown that ALS also exist at the critical point of
the Anderson transition with a finite probability even in infinite
systems.\cite{Obuse1,Obuse2,Obuse3} It is well known that wavefunctions
at metal-insulator transition point show multifractality reflecting the
absence of characteristic lengths in critical
wavefunctions.\cite{Aoki1,Wegner1,Nakayama1} Because of a confining
length of ALS, ALS wavefunctions are not, however, multifractal. The
finite probability of ALS implies that ALS contradict well known
critical properties such as universality, the scaling concept, and the
idea of renormalization. Our results suggest that these critical properties
should be considered for typical critical states. It is natural to
suppose that the critical level statistics is also influenced by ALS.
Our previous works, however, show that ALS do not contribute to the
level statistics.\cite{Obuse3} To understand this fact, it is quite
important to reveal the tail structure of ALS wavefunctions.

In this paper, we investigate spatial structures of ALS wavefunctions
at the critical point of the Anderson transition in
the two-dimensional symplectic class. It is found that the amplitude
distribution of ALS in their tail region actually shows multifractality
and values of exponents characterizing multifractality are the same
with those for typical multifractal wavefunctions for this universality
class. This paper is organized as follows. In \S \ref{sec:2}, we give a
quantitative definition of ALS at the critical point based on the idea
that ALS at criticality do not show multifractality. In \S \ref{sec:3},
the basic multifractal analysis is remained. We briefly explain, in \S
\ref{sec:4}, the employed SU(2) model which belongs to the symplectic class and
a numerical method to obtain eigenstates of systems described by this
model. Numerical results and conclusions are given in \S \ref{sec:5}.

\section{Definition of Anomalously Localized States}
\label{sec:2}

In order to study ALS, we employ a definition of ALS proposed in
ref.~\citen{Obuse1}. This definition is based on the idea that ALS are
not multifractal as a consequence of their localized nature. At first,
we introduce the box-measure correlation function $G_q(l,L,r)$ defined
by\cite{Cates1,Janssen2}
\begin{equation}
G_q (l,L,r) = \frac{1}{N_b N_{b_r}} \sum_b \sum_{b_r} \mu_{b(l)}^q \mu_{b_{r}(l)}^q ,
\label{eq:1}
\end{equation}
where $\mu_{b(l)}=\sum_{i \in b (l)} |\psi_i|^2$ and
$\mu_{b_r(l)}=\sum_{i \in b_r (l)} |\psi_i|^2$ are box measures for
wavefunction amplitudes $\psi_i$, in a box $b(l)$ of size $l$ and in a
box $b_r(l)$ of size $l$ fixed distance $r-l$ away from the box $b(l)$,
respectively. $N_b$ (or $N_{b_r}$) is the number of boxes $b(l)$ [or
$b_r(l)$], and the summation $\sum_{b}$ (or $\sum_{b_r}$) is taken over
all boxes $b(l)$ [or $b_r(l)$] in the system of size $L$. If a
wavefunction is multifractal, $G_q(l,L,r)$ should behave as
\cite{Janssen2}
\begin{equation}
G_q(l,L,r) \propto l^{x(q)} L^{-y(q)} r^{-z(q)},
\label{eq:2}
\end{equation}
where $x(q)$, $y(q)$, and $z(q)$ are exponents describing multifractal
correlations of the amplitude distribution. This relation is sensitive
to ALS as demonstrated in ref.~\citen{Obuse1} and then suitable for
defining ALS. To find the $l$ and $r$ dependences of $G_q(l,L,r)$, we
concentrate on the following functions,
\begin{equation}
Q_q(l) = G_q(l,L,r=l) \propto l^{x(q)-z(q)},
\label{eq:3}
\end{equation}
and
\begin{equation}
R_q(r) = G_q(l=1,L,r) \propto r^{-z(q)}.
\label{eq:4}
\end{equation}
In order to quantify non-multifractality of a specific wavefunction, it
is convenient to introduce variances $\text{Var} (\log Q_2)$ and
$\text{Var} (\log R_2)$ from the linear functions of $\log l$ and $\log
r$, $\log Q_{2}(l) = [x(2)-z(2)] \log l + c_Q$ and $\log R_{2}(r) =
-z(2) \log r + c_R$, respectively, calculated by the least-square fit.
From these variances, we define a quantity $\Gamma$ as
\begin{equation}
\Gamma = \lambda \text{Var}(\log Q_2) + \text{Var} (\log R_2),
\label{eq:5}
\end{equation}
where $\lambda$ is a factor to compensate the difference between
average values of $\text{Var}(\log Q_2)$ and $\text{Var}(\log R_2)$.
Using $\Gamma$ given by eq.~(\ref{eq:5}), the quantitative and
expediential definition of ALS at criticality is presented by
\begin{equation}
\Gamma > \Gamma^*,
\label{eq:6}
\end{equation}
where $\Gamma^*$ is a criterial value of $\Gamma$ to distinguish ALS
from multifractal states.

\section{Multifractal Analysis}
\label{sec:3}

In this section, we give some definitions of basic quantities and
exponents used in the multifractal analysis. At first, we introduce a
quantity $Z(q)$ defined by
\begin{equation}
Z_q (l) = \sum_b \mu_{b (l)}^q.
\label{eq:7}
\end{equation}
Here, symbols and notations in eq.~(\ref{eq:7}) have the same meanings
of those in eq.~(\ref{eq:1}). For a multifractal wavefunction,
$Z_{q}(l)$ obeys a power law
\begin{equation}
Z_q (l) \propto l^{\tau(q)},
\label{eq:8}
\end{equation}
where $\tau(q)$ is called the mass exponent. From eq.~(\ref{eq:8}), the
mass exponent is given by
\begin{equation}
\tau(q) = \lim_{l \rightarrow 0} \frac{\log Z_q(l)}{\log l}.
\label{eq:9}
\end{equation}
Using $\tau(q)$, the generalized dimension $D_q$ is defined as
\begin{equation}
D_q=\frac{\tau(q)}{q-1},
\label{eq:10}
\end{equation}
particulary, $D_q$ for $q=2$ (thus $D_2$) is called the correlation
dimension.

The multifractal spectrum $f(\alpha)$ is defined by the Legendre
transform of $\tau(q)$, i.e., $f(\alpha)=\alpha q - \tau(q)$, where the
Lipschitz-H{\"o}lder exponent $\alpha$ is defined by
$\alpha=d\tau(q)/dq$. The multifractal spectrum $f(\alpha)$ has the
meaning of the fractal dimension of the spatial distribution of boxes
characterized by the exponent $\alpha$. The multifractal spectrum
$f(\alpha)$ and the Lipschitz-H{\"o}lder $\alpha$ can be also
calculated directly from box measures of wavefunction amplitudes as
\cite{Chhabra1}
\begin{equation}
f(\alpha)=\lim_{l\rightarrow 0} \frac{\sum_b m_{b(l)}(q) \log m_{b(l)}(q)}{\log l},
\label{eq:11}
\end{equation}
and
\begin{equation}
\alpha=\lim_{l \rightarrow 0} \frac{\sum_b m_{b(l)}(q) \log \mu_{b(l)}}{\log l},
\label{eq:12}
\end{equation}
respectively. Here, $m_{b(l)}(q)$ called the $q$-microscope is defined
by
\begin{equation}
m_{b(l)}(q)=\frac{\mu_{b(l)}^q}{\sum_{b'} \mu_{b'(l)}^q}.
\label{eq:13}
\end{equation}

\section{System and Numerical Method}
\label{sec:4}

Considering the advantage of system sizes, we focus our attention on
the Anderson transition in two-dimensional electron systems with strong
spin-orbit interactions, in which systems have no spin-rotational
symmetry but have the time-reversal one. Hamiltonians describing these
systems belong to the symplectic class. Among several models belonging
to this universality class, we adopt the SU(2) model because of its
small scaling corrections. The Hamiltonian of the SU(2)
model\cite{Asada1} is given by
\begin{equation}
\boldsymbol{H} = \sum_{i} \varepsilon_i  \boldsymbol{c}_{i}^\dagger  \boldsymbol{c}_{i}
  -  V \sum_{i,j} \boldsymbol{R}_{ij} \boldsymbol{c}_{i}^\dagger \boldsymbol{c}_{j},
\label{eq:14}
\end{equation}
where $ \boldsymbol{c}_{i}^\dagger$ ($ \boldsymbol{c}_{i}$) is the
creation (annihilation) operator acting on a quaternion state vector,
$\boldsymbol{R}_{ij}$ is the quaternion-real hopping matrix element
between the sites $i$ and $j$, and $\varepsilon_i$ denotes the on-site
random potential distributed uniformly in the interval $[-W/2, W/2]$.
(Bold symbols represent quaternion-real quantities.) The matrix element
$\boldsymbol{R}_{ij}$ is given by
\begin{eqnarray}
\boldsymbol{R}_{ij} &=& \cos \alpha_{ij} \cos \beta_{ij} \boldsymbol{\tau}^0
+ \sin \gamma_{ij} \sin \beta_{ij}\boldsymbol{\tau}^1
\nonumber
\\
&-& \cos \gamma_{ij} \sin \beta_{ij}\boldsymbol{\tau}^2
+ \sin \alpha_{ij} \cos \beta_{ij}\boldsymbol{\tau}^3,
\label{eq:15}
\end{eqnarray}
for the nearest neighbor sites $i$ and $j$, and $\boldsymbol{R}_{ij}=0$
for otherwise. Here, $\boldsymbol{\tau}^\mu$ ($\mu=0,1,2,3$) is the
primitive element of quaternions.\cite{Kyrala1} Random quantities
$\alpha_{ij}$ and $\gamma_{ij}$ are distributed uniformly in the range
of $[0,2\pi)$, and $\beta_{ij}$ is distributed according to the
probability density $P(\beta) d \beta = \sin(2\beta) d \beta$ for $0
\le \beta \le \pi/2$. Randomly distritbuted hopping matrix elements
shorten the spin relaxation length which is a dominant irrelevant
length scale. Thus, scaling corrections become very small in the SU(2)
model. It is known that the localization length exponent $\nu$ of this
model is $2.73 \pm 0.02$ and the critical disorder $W_c$ is $5.952V$ at
$E=1.0V$.\cite{Asada1}

Critical wavefunctions of the SU(2) model have been calculated by using
the forced oscillator method (FOM)\cite{Nakayama2} extended to
eigenvalue problems of quaternion-real matrices. Of course, the
Hamiltonian eq.~(\ref{eq:14}) can be represented by complex numbers,
and we can use the usual FOM for complex Hermitian matrices to solve
the eigenvalue problem. The modified FOM for quaternion-real matrices,
however, enables us to calculate eigenvalues and eigenvectors within
about a half of CPU time.\cite{Obuse4} It should be remarked that the
obtained eigenvector is a quaternion-real vector. This vector
represents two physical states simultaneously, which correspond to the
Kramers doublet. Since the amplitude distribution of these degenerate
states are the same, we analyze one of the calculated Kramers doublet.

\section{Results and Conclusions}
\label{sec:5}

\begin{figure}[t]
\begin{center}
\includegraphics[width=8cm]{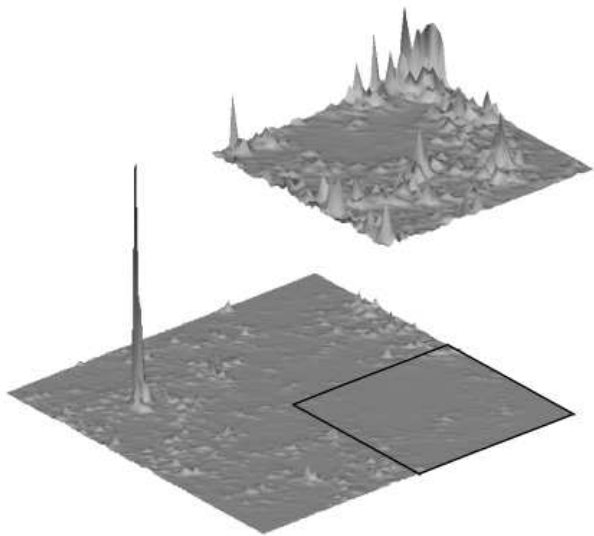}
\end{center}
\caption{Squared amplitudes of a wavefunction $\psi_{\text{ALS}}$ for
the SU(2) model of the system size $L=120$ (lower figure). The value
characterizing non-multifractality defined by eq.~(\ref{eq:5}) is
$\Gamma=0.10$. The square enclosed by solid lines represents a tail
region. The upper figure shows squared amplitudes of the wavefunction
$\psi_{\text{tail}}$ of this tail region (the subsystem size is $L=60$).} \label{fig:1}
\end{figure}

In this section, we show the results of the multifractal analysis for
amplitude distribution of ALS in the tail region. At first, we look for
ALS according to our definition of eq.~(\ref{eq:6}). In this work, the
criterial value is $\Gamma^*=0.03$. For this purpose, we calculate $10^4$
critical wavefunctions for the SU(2) model at $E=1.0V$ and $W=5.952V$ by means of
the FOM. Each eigenstate is obtained for a single disorder realization.
Periodic boundary conditions are imposed in the $x$ and $y$ directions
in the system of size $L=120$.

Figure 1 shows squared amplitudes of a ALS wavefunction $\psi_{\text{ALS}}$.
The value of $\Gamma$ of this
wavefunction is $\Gamma=0.10$. In this wavefunction, amplitudes
concentrate in a narrow spatial region. The wavefunction appears to be
localized for this feature. In order to investigate the tail
structure of this wavefunction, we extract a part of the wavefunction
within a $60 \times 60$ subsystem where the cut region is depicted in
Fig.~\ref{fig:1} by solid lines. The upper figure of Fig.~\ref{fig:1}
shows the extracted tail region wavefunction $\psi_{\text{tail}}$. The
amplitude distribution of $\psi_{\text{tail}}$ is very complicated, and
seems to be multifractal. To clarify this point, we examine below
multifractality of $\psi_{\text{tail}}$.

\begin{figure}[t]
\begin{center}
\includegraphics[width=8cm]{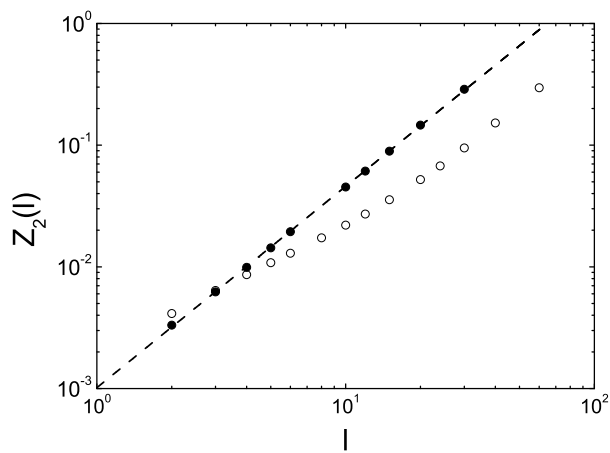}
\end{center}
\caption{$Z_2(l)$ for $\psi_{\text{ALS}}$ (open circles) and
$\psi_{\text{tail}}$ (filled circles). The dashed line shows the least
square fit for $Z_2(l)$ for $\psi_{\text{tail}}$.} \label{fig:2}
\end{figure}

At first, we calculate $Z_2(l)$ given by eq.~(\ref{eq:7}) for
$\psi_{\text{ALS}}$ and $\psi_{\text{tail}}$. The wavefunction
$\psi_{\text{tail}}$ is normalized in the subsystem to perform the
multifractal analysis. We see that
$Z_2(l)$ for $\psi_{\text{ALS}}$ does not exhibit the power law as
shown in Fig.~\ref{fig:2}  (open circles). If ALS are truly localized
such as $\psi_{\text{ALS}}(r) \propto \exp(-r/\xi)$, where $\xi$ is a
localization length, $Z_2(l)$ for $\psi_{\text{ALS}}$ should take a
constant value for $l \gg \xi$. However, $Z_2(l)$ for ALS is not
constant even for large $l$ as shown in Fig.~\ref{fig:2}. On the contrary,
the quantity $Z_2(l)$ for $\psi_{\text{tail}}$ clearly obeys a power
law as shown by filled circles in Fig.~\ref{fig:2}. The value of the
correlation dimension $D_2=\tau(2)=1.69\pm0.01$ is very close to the
value reported so far for typical multifractal wavefunctions belonging
to this universality class.\cite{Obuse1} This implies that tail regions
of ALS exhibit the same multifractality with that of typical critical
wavefunctions. The fact that the behavior of $Z_2(l)$ for
$\psi_{\text{ALS}}$ for $l \gg 20$ is similar to that for
$\psi_{\text{tail}}$ shows that $Z_2(l)$ at large $l$ values is
dominated by the tail structure of the ALS wavefunction.

\begin{figure}[b]
\begin{center}
\includegraphics[width=8cm]{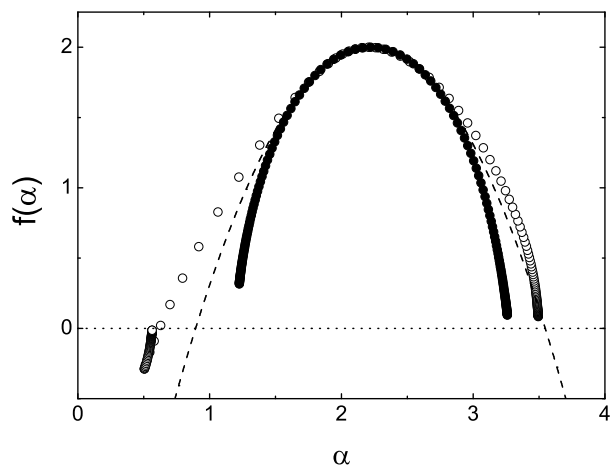}
\end{center}
\caption{Multifractal spectra $f(\alpha)$ for $\psi_{\text{ALS}}$ (open
circles) and $\psi_{\text{tail}}$ (filled circles). The dashed line
shows the parabolic approximation eq.~(\ref{eq:16}) with
$\alpha_0=2.15$.} \label{fig:3}
\end{figure}

Next, we calculate the multifractal spectra for both wavefunctions
$\psi_{\text{ALS}}$ and $\psi_{\text{tail}}$. Results are shown in
Fig.~\ref{fig:3}. Remarkably, $f(\alpha)$ for $\psi_{\text{ALS}}$ takes
negative values for small $\alpha$, which is not reasonable because
$f(\alpha)$ is the fractal dimension of the distribution of boxes
characterized by $\alpha$. This supports that $\psi_{\text{ALS}}$ does
not have a multifractal structure. On the other hand, $f(\alpha)$ for
$\psi_{\text{tail}}$ satisfies several conditions for multifractal
spectra. The value of $\alpha_0$ giving the maximum of $f(\alpha)$ for
$\psi_{\text{tail}}$ is $\alpha_0=2.15\pm0.02$, and the minimum and
maximum values of $\alpha$ are estimated as $\alpha_{\text{min}}=1.22$
and $\alpha_{\text{max}}=3.26$, respectively.
These values are very close to those for typical multifractal wavefunctions.
In fact, we calculate $f(\alpha)$ for a typical critical wavefunction
with $\Gamma=0.001$ which is the smallest value of $\Gamma$ in all
calculated wavefunctions. Obtained values are $\alpha_0=2.17$,
$\alpha_{\text{min}}=1.22$, and $\alpha_{\text{max}}=3.31$. The
characteristic values of $\alpha$ for $\psi_{\text{tail}}$ are also
close to averaged values for an ensemble of typical critical
wavefunctions.\cite{Obuse2}
The profile of
$f(\alpha)$ for $\psi_{\text{tail}}$ in the vicinity of
$\alpha=\alpha_0$ can be well approximated by the parabolic form
\begin{equation}
f(\alpha)=2-\frac{(\alpha_0-\alpha)^2}{4(\alpha_0-2)},
\label{eq:16}
\end{equation}
which is shown by dashed line in Fig.~\ref{fig:3}.
These results ensure that $\psi_{\text{tail}}$ has the same multifractal
amplitude distribution with that for typical critical wavefunctions.
It seems that
$f(\alpha)$ for $\psi_{\text{ALS}}$ is also fitted by
eq.~(\ref{eq:16}). This possibility is, however, denied when examining
the generalized dimension $D_q$.

\begin{figure}[t]
\begin{center}
\includegraphics[width=8cm]{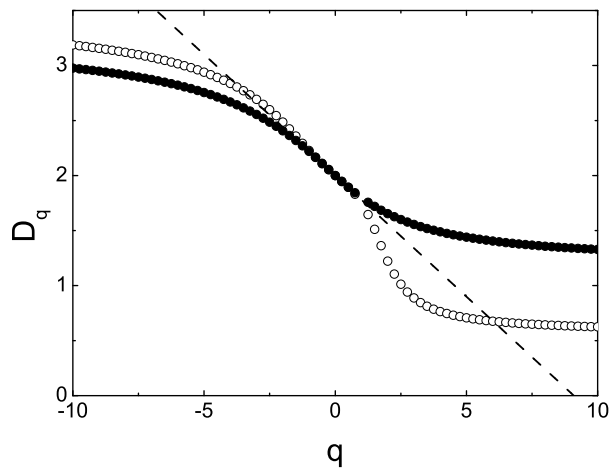}
\end{center}
\caption{The generalized dimension as a function of the moment $q$ for
$\psi_{\text{ALS}}$ (open circles) and $\psi_{\text{tail}}$ (filled
circles). The dashed line represents the parabolic approximation
eq.~(\ref{eq:17}) with $\alpha_0=2.15$.} \label{fig:4}
\end{figure}

Figure \ref{fig:4} shows the generalized dimension $D_q$ calculated by
the box-counting method eq.~(\ref{eq:7}) for both wavefunctions
$\psi_{\text{ALS}}$ and $\psi_{\text{tail}}$. The values of $D_q$ for
two wavefunctions largely deviate each other for $|q|\gg1$. Since
$D_{-\infty}$ and $D_\infty$ are equal to $\alpha_{\text{max}}$ and
$\alpha_{\text{min}}$, respectively, such large deviations are
consistent with Fig.~\ref{fig:3}. The parabolic approximation of $D_q$
corresponding to eq.~(\ref{eq:16}) is given by
\begin{equation}
D_q=2-q(\alpha_0-2),
\label{eq:17}
\end{equation}
which would be valid in the vicinity of $q=0$. The generalized
dimension $D_q$ for $\psi_{\text{tail}}$ is well approximated by
eq.~(\ref{eq:17}) with $\alpha_0=2.15$ as shown by dashed line in
Fig.~\ref{fig:4}. On the contrary, $D_q$ for $\psi_{\text{ALS}}$ does
not follow a straight line near $q=0$. This implies that the parabolic
approximation eq.~(\ref{eq:16}) does not reproduce $f(\alpha)$ for
$\psi_{\text{ALS}}$ even in the vicinity of $\alpha=\alpha_0$.

In summary, we investigate the spatial structure of ALS at the critical
point of the Anderson transition. It is found that ALS are not truly
localized and the tail structure of ALS shows the same multifractality
with that of typical wavefunctions. These
results form the foundation of understanding the distribution of
physical quantities at criticality.

\begin{acknowledgements}
We are grateful to T. Nakayama for helpful discussions. This work was
supported in part by a Grant-in-Aid for Scientific Research from Japan
Society for the Promotion of Science (No.~$14540317$) and
for the 21st Century Centre of Excellence (COE) Program, entitled
``Topological Science and Technology'', from the Ministry of Education,
Culture, Sport, Science and Technology of Japan (MECSST). Numerical
calculations in this work have been mainly performed on the facilities
of the Supercomputer Center, Institute for Solid State Physics,
University of Tokyo.
\end{acknowledgements}


\begin{thebibliography}{99}
\bibitem{Obuse4}
Details of the forced oscillator method for quaternion-real
matrices will be published elsewhere.
\end{thebibliography}
\end{document}